\documentclass{amsart}

\usepackage{amsmath,amssymb,amsfonts,amscd,amsthm,amsopn}  
\usepackage{amsthm}
\usepackage{amssymb}
\usepackage{latexsym}
\usepackage[notref,notcite]{showkeys} 

\usepackage{hyperref}
\hypersetup{
    citecolor=blue,        
}

\usepackage{color}

\newtheorem{prop}{Proposition}[section]

\newtheorem{defn}{Definition}[section]

\newtheorem{rmk}{Remark}

\numberwithin{equation}{section}
\newcommand{\beqa}{\begin{eqnarray}}
\newcommand{\eeqa}{\end{eqnarray}}

\newcommand{\nc}{\newcommand}
\newcommand{\rnc}{\renewcommand}

\nc{\cal}{\mathcal}

\nc{\goth}{\mathfrak}
\rnc{\bold}{\mathbf}
\renewcommand{\frak}{\mathfrak}
\renewcommand{\Bbb}{\mathbb}

\nc{\Cal}{\mathcal}
\nc{\Xp}[1]{X^+(#1)}
\nc{\Xm}[1]{X^-(#1)}
\nc{\on}{\operatorname}
\nc{\ch}{\mbox{ch}}
\nc{\Z}{{\bold Z}}
\nc{\J}{{\mathcal J}}
\nc{\C}{{\bold C}}
\nc{\Q}{{\bold Q}}
\nc{\oC}{{\widetilde{C}}}
\nc{\oc}{{\tilde{c}}}
\nc{\og}{{\tilde{\gamma}}}
\nc{\lC}{{\overline{C}}}
\nc{\lc}{{\overline{c}}}
\nc{\Rt}{{\tilde{R}}}

\nc{\odel}{{\overline{\delta}}}

\nc{\N}{{\Bbb N}}
\nc\beq{\begin{equation}}
\nc\enq{\end{equation}}
\nc\lan{\langle}
\nc\ran{\rangle}
\nc\bsl{\backslash}
\nc\mto{\mapsto}
\nc\lra{\leftrightarrow}
\nc\hra{\hookrightarrow}
\nc\sm{\smallmatrix}
\nc\esm{\endsmallmatrix}
\nc\sub{\subset}
\nc\ti{\tilde}
\nc\nl{\newline}
\nc\fra{\frac}
\nc\und{\underline}
\nc\ov{\overline}
\nc\ot{\otimes}
\nc\bbq{\bar{\bq}_l}
\nc\bcc{\thickfracwithdelims[]\thickness0}
\nc\ad{\text{\rm ad}}
\nc\Ad{\text{\rm Ad}}
\nc\Hom{\text{\rm Hom}}
\nc\End{\text{\rm End}}
\nc\Ind{\text{\rm Ind}}
\nc\Res{\text{\rm Res}}
\nc\Ker{\text{\rm Ker}}
\rnc\Im{\text{Im}}
\nc\sgn{\text{\rm sgn}}
\nc\tr{\text{\rm tr}}
\nc\Tr{\text{\rm Tr}}
\nc\supp{\text{\rm supp}}
\nc\card{\text{\rm card}}
\nc\bst{{}^\bigstar\!}
\nc\he{\heartsuit}
\nc\clu{\clubsuit}
\nc\spa{\spadesuit}
\nc\di{\diamond}
\nc\cW{\cal W}
\nc\cG{\cal G}
\nc\al{\alpha}
\nc\bet{\beta}
\nc\ga{\gamma}
\nc\de{\delta}
\nc\ep{\epsilon}
\nc\io{\iota}
\nc\om{\omega}
\nc\si{\sigma}
\rnc\th{\theta}
\nc\ka{\kappa}
\nc\la{\lambda}
\nc\ze{\zeta}

\nc\vp{\varpi}
\nc\vt{\vartheta}
\nc\vr{\varrho}

\nc\Ga{\Gamma}
\nc\De{\Delta}
\nc\Om{\Omega}
\nc\Si{\Sigma}
\nc\Th{\Theta}
\nc\La{\Lambda}

\nc\boa{\bold a}
\nc\bob{\bold b}
\nc\boc{\bold c}
\nc\bod{\bold d}
\nc\boe{\bold e}
\nc\bof{\bold f}
\nc\bog{\bold g}
\nc\boh{\bold h}
\nc\boi{\bold i}
\nc\boj{\bold j}
\nc\bok{\bold k}
\nc\bol{\bold l}
\nc\bom{\bold m}
\nc\bon{\bold n}
\nc\boo{\bold o}
\nc\bop{\bold p}
\nc\boq{\bold q}
\nc\bor{\bold r}
\nc\bos{\bold s}
\nc\bou{\bold u}
\nc\bov{\bold v}
\nc\bow{\bold w}
\nc\boz{\bold z}

\nc\ba{\bold A}
\nc\bb{\bold B}
\nc\bc{\bold C}
\nc\bd{\bold D}
\nc\be{\bold E}
\nc\bg{\bold G}
\nc\bh{\bold H}
\nc\bi{\bold I}
\nc\bj{\bold J}
\nc\bk{\bold K}
\nc\bl{\bold L}
\nc\bm{\bold M}
\nc\bn{\bold N}
\nc\bo{\bold O}
\nc\bp{\bold P}
\nc\bq{\bold Q}
\nc\br{\bold R}
\nc\bs{\bold S}
\nc\bt{\bold T}
\nc\bu{\bold U}
\nc\bv{\bold V}
\nc\bw{\bold W}
\nc\bz{\bold Z}
\nc\bx{\bold X}

\nc\ca{\mathcal A}
\nc\cb{\mathcal B}
\nc\cc{\mathcal C}
\nc\cd{\mathcal D}
\nc\ce{\mathcal E}
\nc\cf{\mathcal F}
\nc\cg{\mathcal G}
\rnc\ch{\mathcal H}
\nc\ci{\mathcal I}
\nc\cj{\mathcal J}
\nc\ck{\mathcal K}
\nc\cl{\mathcal L}
\nc\cm{\mathcal M}
\nc\cn{\mathcal N}
\nc\co{\mathcal O}
\nc\cp{\mathcal P}
\nc\cq{\mathcal Q}
\nc\car{\mathcal R}
\nc\cs{\mathcal S}
\nc\ct{\mathcal T}
\nc\cu{\mathcal U}
\nc\cv{\mathcal V}
\nc\cz{\mathcal Z}
\nc\cx{\mathcal X}
\nc\cy{\mathcal Y}

\nc\e[1]{E_{#1}}
\nc\ei[1]{E_{\delta - \alpha_{#1}}}
\nc\esi[1]{E_{s \delta - \alpha_{#1}}}
\nc\eri[1]{E_{r \delta - \alpha_{#1}}}
\nc\ed[2][]{E_{#1 \delta,#2}}
\nc\ekd[1]{E_{k \delta,#1}}
\nc\emd[1]{E_{m \delta,#1}}
\nc\erd[1]{E_{r \delta,#1}}

\nc\ef[1]{F_{#1}}
\nc\efi[1]{F_{\delta - \alpha_{#1}}}
\nc\efsi[1]{F_{s \delta - \alpha_{#1}}}
\nc\efri[1]{F_{r \delta - \alpha_{#1}}}
\nc\efd[2][]{F_{#1 \delta,#2}}
\nc\efkd[1]{F_{k \delta,#1}}
\nc\efmd[1]{F_{m \delta,#1}}
\nc\efrd[1]{F_{r \delta,#1}}

\def\sA{{\textsf A}}

\nc\fa{\frak a}
\nc\fb{\frak b}
\nc\fc{\frak c}
\nc\fd{\frak d}
\nc\fe{\frak e}
\nc\ff{\frak f}
\nc\fg{\frak g}
\nc\fh{\frak h}
\nc\fj{\frak j}
\nc\fk{\frak k}
\nc\fl{\frak l}
\nc\fm{\frak m}
\nc\fn{\frak n}
\nc\fo{\frak o}
\nc\fp{\frak p}
\nc\fq{\frak q}
\nc\fr{\frak r}
\nc\fs{\frak s}
\nc\ft{\frak t}
\nc\fu{\frak u}
\nc\fv{\frak v}
\nc\fz{\frak z}
\nc\fx{\frak x}
\nc\fy{\frak y}

\nc\cO{\cal O}
\nc\cU{\cal U}
\nc\cA{\cal A}
\nc\cB{\cal B}
\nc\cK{\cal K}
\nc\cR{\cal R}

\nc\fA{\frak A}
\nc\fB{\frak B}
\nc\fC{\frak C}
\nc\fD{\frak D}
\nc\fE{\frak E}
\nc\fF{\frak F}
\nc\fG{\frak G}
\nc\fH{\frak H}
\nc\fJ{\frak J}
\nc\fK{\frak K}
\nc\fL{\frak L}
\nc\fM{\frak M}
\nc\fN{\frak N}
\nc\fO{\frak O}
\nc\fP{\frak P}
\nc\fQ{\frak Q}
\nc\fR{\frak R}
\nc\fS{\frak S}
\nc\fT{\frak T}
\nc\fU{\frak U}
\nc\fV{\frak V}
\nc\fZ{\frak Z}
\nc\fX{\frak X}
\nc\fY{\frak Y}
\nc\tfi{\ti{\Phi}}
\nc\bF{\bold F}
\rnc\bol{\bold 1}

\newcommand{\hs}[1]{\hspace{#1 mm}}
\newcommand{\mb}[1]{\hs{4}\mbox{#1}\hs{4}}

\nc\ua{\bold U_\A}

\nc\RR{\mathbb R}
\nc\CC{\mathbb C}
\nc\II{\mathbb I}
\nc\qinti[1]{[#1]_i}
\nc\q[1]{[#1]_q}
\nc\xpm[2]{E_{#2 \delta \pm \alpha_#1}}  
\nc\xmp[2]{E_{#2 \delta \mp \alpha_#1}}
\nc\xp[2]{E_{#2 \delta + \alpha_{#1}}}
\nc\xm[2]{E_{#2 \delta - \alpha_{#1}}}
\nc\hik{\ed{k}{i}}
\nc\hjl{\ed{l}{j}}
\nc\qcoeff[3]{\left[ \begin{smallmatrix} {#1}& \\ {#2}& \end{smallmatrix}
\negthickspace \right]_{#3}}
\nc\qi{q}
\nc\qj{q}

\nc\ufdm{{_\ca\bu}_{\rm fd}^{\le 0}}

\nc\nonu{\nonumber\\}

\nc\isom{\cong} 

\nc{\pone}{{\Bbb C}{\Bbb P}^1}
\nc{\pa}{\partial}

\nc{\F}{{\mathcal F}}
\nc{\Sym}{{\goth S}}
\nc{\A}{{\mathcal A}}
\nc{\arr}{\rightarrow}
\nc{\larr}{\longrightarrow}

\nc{\ri}{\rangle}
\nc{\lef}{\langle}
\nc{\W}{{\mathcal W}}
\nc{\uqatwoatone}{{U_{q,1}}(\su)}
\nc{\uqtwo}{U_q(\goth{sl}_2)}
\nc{\dij}{\delta_{ij}}
\nc{\divei}{E_{\alpha_i}^{(n)}}
\nc{\divfi}{F_{\alpha_i}^{(n)}}
\nc{\Lzero}{\Lambda_0}
\nc{\Lone}{\Lambda_1}
\nc{\ve}{\varepsilon}
\nc{\phioneminusi}{\Phi^{(1-i,i)}}
\nc{\phioneminusistar}{\Phi^{* (1-i,i)}}
\nc{\phii}{\Phi^{(i,1-i)}}
\nc{\Li}{\Lambda_i}
\nc{\Loneminusi}{\Lambda_{1-i}}
\nc{\vtimesz}{v_\ve \otimes z^m}

\nc{\asltwo}{\widehat{\goth{sl}_2}}
\nc\ag{\widehat{\goth{g}}}  
\nc\teb{\tilde E_\boc}
\nc\tebp{\tilde E_{\boc'}}

\newcommand{\ben}{\begin{eqnarray}}
\newcommand{\een}{\end{eqnarray}}

\begin{document}

\title{Generalized $q$-Onsager algebras and dynamical $K$-matrices}
\author{S. Belliard}
\address{Istituto Nazionale di Fisica Nucleare, Sezione di Bologna, Via Irnerio 46,  40126 Bologna,  Italy}
\email{belliard@bo.infn.it}
\author{V. Fomin}
\address{
Laboratoire de Physique Th\'eorique LAPTH
Universit\'e de Savoie Ð CNRS (UMR 5108) BP 110, F-74941 Annecy-le-Vieux Cedex, France.
}
\email{fomin@lapp.in2p3.fr}

\begin{abstract}
A procedure to construct $K$-matrices from the generalized $q$-Onsager algebra $\cO_{q}(\widehat{g})$ is proposed. This procedure extends the intertwiner techniques used to obtain scalar (c-number) solutions of the reflection equation to dynamical (non-c-number) solutions. It shows the relation between soliton non-preserving reflection equations or twisted reflection equations and the generalized $q$-Onsager algebras. These dynamical $K$-matrices are important to quantum integrable models with extra degrees of freedom located at the boundaries: for instance, in the quantum affine Toda field theories on the half-line they yield the boundary amplitudes. As examples, the cases of $\cO_{q}(a^{(2)}_{2})$ and $\cO_{q}(a^{(1)}_{2})$ are treated in details. 
\end{abstract}

\maketitle

\vskip -0.2cm

{{\small  {\it \bf Keywords}: generalized  $q$-Onsager algebra; reflection algebras; intertwiner equation; Zhedanov algebra; dynamical boundary conditions.}}


\section{Introduction}

\vspace{0.5cm}

Mathematical structure behind the Yang-Baxter equation (YBE), namely quantum groups, gives a simple way to obtain the $R$-matrix as a solution of the intertwiner equation.
For instance, the quantum affine algebras $\cU_{q}(\widehat{g})$, formulated by Drinfel'd and Jimbo \cite{D1,J1}, allow to construct trigonometric $R$-matrices \cite{J}, solutions of the Yang-Baxter equation. 
These algebras admit different formulations: 1) $q$-Serre-Chevalley formulation\footnote{in the limit $q \to 1$ one recovers the Serre-Chevalley formulation of the affine Lie algebras.} \cite{D1,J1}; 2) FRT formulation \cite{FRT,RS} which appeared in the context of the Quantum Inverse Scattering Method (QISM) for quantum integrable systems; 3) Drinfel'd second realization \cite{D2}. Each of these formulations has shown its interest in different physical and mathematical problems : the $q$-Serre-Chevalley formulation of quantum algebra leads to the construction of the $R$-matrices \cite{J} and its realization in terms of field operators corresponds to non-local conserved currents of the affine Toda field theories (ATFT) \cite{BL}; the FRT formulation gives a simple construction of abelian conserved quantities of bulk integrable models and in many cases it allows the diagonalization of the transfer matrix using the so-called algebraic Bethe ansatz technique \cite{STF}; the Drinfel'd second realization leads to infinite dimensional representations  of the algebra \cite{FJ}, the calculation of correlation functions using vertex operators techniques \cite{Jall} or the calculation of scalar products of Bethe vectors \cite{BPR}.      
   
\vspace{0.2cm}

In the case of quantum integrable models with boundaries, the reflection equations that ensure the integrability of these models are known since seminal works of Cherednik and Sklyanin \cite{Ch,Sk}.  The related algebraic structures are given by coideal subalgebras of the quantum algebras. 
These subalgebras are called reflection algebras and the commutation relations are encoded by the reflection equations \cite{MRS}.
The definition of the reflection algebras can be constructed from the FRT formulation of the quantum algebra $\cU_q(\widehat{g})$ using two different automorphisms. 
The first one is related to the inverse the monodromy matrix and we call it the reflection algebra (RA). The second one is related to the transposition of the monodromy matrix and we call it the Twisted reflection algebra (TRA). In both cases the algebras are given in a FRT type formulation and the question of a corresponding $q$-Serre-Chevalley or Drinfel'd second realization type formulation is mostly open. For the finite Lie algebra $g$ the $q$-Serre-Chevalley type formulation of the reflection equation is known and appears in the context of the quantum symmetric pairs and related q-orthogonal polynomials \cite{NS,Le}. For special choices of $g$ it corresponds to special cases of the Zhedanov or Askey-Wilson algebra \cite{Zhed} related to the Askey-Wilson polynomials. Moreover, in the case $\cU_q(a^{(1)}_1)$ an isomorphism of coideal subalgebra has been shown between the reflection algebra and the $q$-Onsager algebra $\cO_{q}(a^{(1)}_1)$ (which correspond to a $q$-Serre-Chevalley formulation of the reflection equation) \cite{BasS,BB2}.

\vspace{0.2cm}

In a recent work \cite{BB}, the generalized $q$-Onsager algebra $\cO_{q}(\widehat{g})$ was introduced as a coideal subalgebra of the quantum algebra $\cU_{q}(\widehat{g})$. Similarly to the calculation of the $R$-matrices using the $q$-Serre-Chevalley formulation of $\cU_{q}(\widehat{g})$ and the intertwiner equation \cite{J1}, the generalized $q$-Onsager algebra $\cO_{q}(\widehat{g})$ allows the calculation of the $K$-matrices, solutions of the twisted reflection equation. These $\cO_{q}(\widehat{g})$ algebras emerge as the closure relations of the non-local charges of the ATFT on the half-line with soliton non-preserving boundary conditions\footnote{corresponding to the Twisted reflection equation.}. 
Note that these non-local charges have been previously constructed from the $\cU_{q}(\widehat{g})$ generators and were sufficient to construct scalar (c-number) boundary scattering amplitudes from the intertwiner equation \cite{MN, DM, DelG} although their closure relations were unknown. 
However, the identification of the $\cO_{q}(\widehat{g})$ algebra as a non-abelian symmetry of the ATFT on the half-line allows to proceed further: it gives the complete description of the admissible soliton non-preserving boundary conditions: 1) scalar (c-number) ones, which reproduce the known results \cite{Call} and 2) dynamical (non-c-number) ones \cite{BB}. 
As a consequence, the knowledge of $\cO_{q}(\widehat{g})$ allows to extend the intertwiner technique to dynamical (non-c-number) $K$-matrices in a systematic way.

\vspace{0.2cm}

From a physical point of view, the dynamical boundary conditions correspond to quantum integrable models with degrees of freedom at the boundary. Some examples are given by conformal models perturbed by a dynamical boundary as the massless boundary Sine-Gordon model related to the quantum impurity problem \cite{Sal},  or the generalisation to massless boundary ATFT $a^{(1)}_2$ \cite{BHK}. 
Extensions to the massive boundary Sine-Gordon model have also been considered \cite{BassL,BasD,BK}. For one of them  \cite{BK}, a dynamical $K$-matrix has been obtained using intertwiner technique. Among the properties of this model, it is important to notice that for real coupling the boundary amplitudes are self-dual \cite{BK}, as for the bulk case \cite{Call2}, contrary to the case with scalar boundary conditions \cite{G,Cor}.  
There are also examples of dynamical boundary conditions in quantum integrable spin chains, where the dynamical $K$-matrices are obtained as 'dressing' of scalar (c-number) boundary matrices \cite{DynBHubb,DynBtJ,FSlav}.

\vspace{0.2cm}

In this paper we present dynamical (non-c-number) $K$-matrices obtained using the intertwiner equation associated with the generalized $q$-Onsager algebras for two cases :  $\cO_q(a^{(2)}_2)$ and $\cO_q(a^{(1)}_2)$. These two examples show an explicit connection between the generalized $q$-Onsager algebras and the reflection algebras. 
In Section 2 we recall some basics about the quantum algebras, reflection algebras, intertwiner equations and explain the construction of the dynamical $K$-matrices.
In Section 3 we give the $K$-matrices for  $\cO_q(a^{(2)}_2)$ and $\cO_q(a^{(1)}_2)$ realized in terms of the
Zhedanov algebra \cite{Zhed}. {\ The mapping} to scalar (c-number) matrices is described in Section 4 recovering known results. 
Extensions to other affine Lie algebras, applications in physics and further investigations are discussed in Section 5.

\vspace{0.5cm}


\section{Affine quantum algebras, reflection equations and generalized $q$-Onsager algebras}

In this section we briefly recall basic notions on the affine quantum algebras and their coideals. We emphasize the similarities between the FRT formalism (RLL relation) and intertwiner equation (quasi-triangular Hopf structure) for quantum algebras \cite{Chari} and the reflection equation and intertwiner equation for coideals of quantum algebras \cite{MN,DM,N}.

We point out that generalized $q$-Onsager algebras are realized as coideal subalgebras of quantum algebras and the knowledge of these algebras both with the intertwiner equation provides a simple method for the derivation of solutions of the reflection equation. In the end of the section we give a strategy of the construction of dynamical (non-c-number or operator-valued) $K$-matrices. 

\vspace{0.2cm}
\subsection{Hopf algebra and the Yang-Baxter equation.} 
In mathematics, the affine quantum algebras $\cU_{q}(\widehat{g})$ are quasi-triangular Hopf algebras (see \cite{Chari} for details). Let us consider the coproduct map $\Delta : \cU_{q}(\widehat{g}) \to \cU_{q}(\widehat{g}) \otimes \cU_{q}(\widehat{g})$ and the universal $R$-matrix $\cR \in \cU_{q}(\widehat{g})\otimes \cU_{q}(\widehat{g})$.
The universal $R$-matrix is an invertible element such that:
\ben
\cR \,\Delta(x) = \sigma \circ \Delta (x) \, \cR, 
\een
with $x \in \cU_{q}(\widehat{g})$ and $\sigma$ the permutation map defined by $\sigma(a\otimes b)=b\otimes a$. As a consequence of the quasi-triangular properties, the universal $R$-matrix is a solution of the universal Yang-Baxter equation:
\ben
\cR_{12} \cR_{13} \cR_{23}=\cR_{23}\cR_{13}\cR_{12}.  
\een
Using a finite dimensional representation $\pi_u: \cU_{q}(\widehat{g})\to End(V)$, with $V$ a finite dimensional vector space, one maps the universal object to a matrix. It follows that the $R$-matrix satisfies
the intertwiner equation:
\ben
R_{12}(u/v)( \pi_u \otimes \pi_v )[\Delta(x)]= ( \pi_u \otimes \pi_v )[\sigma \circ \Delta(x)]R_{12}(u/v), \quad x \in \cU_{q}(\widehat{g}) \label{IE}
\een
and the Yang-Baxter equation:
\ben
R_{12}(u/v)R_{13}(u)R_{23}(v)=R_{23}(v)R_{13}(u)R_{12}(u/v).\label{YB}
\een
where
\ben
R_{12}(u/v)=( \pi_u \otimes \pi_v )\cR_{12}.
\een
For formal parameters $u$ and $v$, the tensor product of representations $ \pi_u \otimes \pi_v $ is an irreducible representation which implies that the solution of (\ref{IE}) is unique up to a scalar function and also a solution of (\ref{YB})  \cite{J}. These properties allow the construction of the scalar $R$-matrices solution of the Yang-Baxter equation from the representation theory and the coproduct of  $\cU_{q}(\widehat{g})$. 

\vspace{0.2cm}

If one evaluates only the first space of the universal $R$-matrix with $\pi_u$ , one obtains the affine monodromy matrix $L(u)$ \footnote{For simplicity we consider only one family of monodromy matrices, a complete description of the algebra needs two families, see \cite{RS} for details.}:
\ben
L_1(u)=(\pi_u \otimes 1) \cR_{12}. 
\een
Evaluating the spaces 1 and 2 of the universal Yang-Baxter equation one obtains the FRT formulation of  $\cU_{q}(\widehat{g})$:  
\ben
R_{12}(u/v)L_1(u)L_2(v)=L_2(v)L_1(u)R_{12}(u/v).
\een
When $\widehat{g}=a^{(1)}_n$ an evaluation homomorphism $ \Pi:  \cU_q(a^{(1)}_n) \to\cU_q(gl_{n+1})$ has been constructed \cite{J1} and the corresponding finite monodromy matrix obtained from the intertwiner equation:
\ben
L^{\Pi}(u) (\pi_u \otimes \Pi)\Delta (x)=(\pi_u \otimes \Pi)(\sigma \circ \Delta(x)) L^{\Pi}(u). \label{intPiL}
\een
The aim of this paper is to provide the construction of a similar object in the case of the reflection algebras.

\vspace{0.2cm}
\subsection{Reflection algebras and coideal subalgebras.}
The reflection equations are representations of the reflection algebras which can be realized as coideal subalgebras of the quantum affine algebras $\cU_q(\widehat{g})$. Construction of these reflection algebras is related to automorphisms of the FRT algebra \cite{Sk,MRS}. There are two automorphisms of the FRT algebra leading to two different reflection algebras:

\vspace{0.1cm}
 
- The first family, the so-called reflection algebra (RA), noted $\cB(\widehat{g})$, is generated by  $\cK(u) \in End(V)\otimes \cB(\widehat{g}) [[u,u^{-1}]]$ with the commutation relation:
 \ben
R_{12}(u/v)\cK_1(u)R_{21}(uv)\cK_2(v)=\cK_2(v)R_{12}(uv)\cK_1(u)R_{21}(u/v) \label{RA}. 
\een
 The coideal properties follow from the existence of a coaction map $ \delta : \cB(\widehat{g}) \to \cU_q(\widehat{g}) \otimes \cB(\widehat{g}) $ given by :
\ben
\delta( \cK(u)) =L(u)\cK(u)L(u^{-1})^{-1}
\een
and a counit $\epsilon:  \cB(\widehat{g}) \to \CC$ given by:
\ben
\epsilon( \cK(u))=K(u)
\een 
where the matrix $K(u)$ is a scalar solution of (\ref{RA}).
These coideal properties imply that the reflection algebra can be constructed as a subalgebra of  $\cU_q(\widehat{g})$ from the homomorphism $(1\otimes \epsilon)\circ \delta \equiv \psi  : \cB(\widehat{g}) \to\cU_q(\widehat{g})$: 
\ben
\psi(\cK(u))=L(u)K(u)L(u^{-1})^{-1}.  \label{dressRA}
\een
The map $w_1: L(u) \to L^{-1}(u^{-1})$ is an automorphism of the FRT algebra. 

Although a clear algebraic framework for these coideal subalgebras and an 'universal' $K$-matrix is still an open question, at the repesentation level an intertwiner equation can  be also defined for the $K$-matrix  \cite{MN,DM,N}. Assuming that a $q$-Serre-Chevalley formulation of  $\cB(\widehat{g})$ is known we can define the equation :
\ben 
K(u)( \pi_u \otimes \Pi ) \delta(y)=( \pi_{u^{-1}} \otimes \Pi) \delta(y)K(u) , \quad y \in \cB(\widehat{g}) \label{intRA}
\een
with $\Pi$ an evaluation homomorphism to a finite algebra. Using commuting diagrams technique \cite{DM}, one can show that the solution of this equation will satisfy the equation (\ref{RA}). 

\vspace{0.1cm}

- The second family, the so-called twisted reflection algebra (TRA), noted $\cB^*(\widehat{g})$, is generated by  $\cK^*(u) \in End(V)\otimes \cB^*(\widehat{g}) [[u,u^{-1}]]$ with the commutation relation:
\ben
R_{12}(u/v)\cK^*_1(u)R^{t_1}_{12}(1/uv)\cK^*_2(v)=\cK^*_2(v)R^{t_1}_{12}(1/uv)\cK^*_1(u)R_{12}(u/v) \label{TRA}
\een
 The coideal properties follow from the existence of a coaction map $ \delta : \cB^*(\widehat{g}) \to \cU_q(\widehat{g}) \otimes \cB^*(\widehat{g}) $ given by :
\ben
\delta( \cK^*(u)) =L(u)\cK^*(u)L(u^{-1})^{t}
\een
and a counit $\epsilon:  \cB(\widehat{g}) \to \CC$ given by:
\ben
\epsilon( \cK^*(u))=K^*(u)
\een 
where the matrix $K^*(u)$ is a scalar solution of (\ref{TRA}).
These coideal properties imply that the reflection algebra can be constructed as a subalgebra of  $\cU_q(\widehat{g})$ from the homomorphism $(1\otimes \epsilon)\circ \delta=\psi:  \cB^*(\widehat{g}) \to \cU_q(\widehat{g})$: 
\ben
\psi(\cK^*(u))=L(u)K^*(u)L(u^{-1})^{t}. \label{dressTRA}
\een
The map $w_2: L(u) \to L(u^{-1})^{t}$ is an automorphism of the FRT algebra. 

As for the RA case, assuming that a $q$-Serre-Chevalley formulation of  $\cB^*(\widehat{g})$ is known, we can define the intertwiner equation :
\ben
K^*(u)( \pi_u \otimes \Pi ) \delta(y)=( \bar{\pi}_{u^{-1}} \otimes \Pi) \delta(y)K^*(u) , \quad y \in \cB^*(\widehat{g})  \label{intTRA}
\een
with $\Pi$ an evaluation homomorphism to a finite algebra and $\bar{\pi}$ a finite dimensional conjugate representation of $\cU_q(\widehat{g})$ of the same dimension as $\pi$.
Using commuting diagrams \cite{DM}, one can show that the solution of this equation will satisfy the equation (\ref{TRA}).

\vspace{0.1cm}

Let us mention that if the $R$-matrix of $\cU_q(\widehat{g})$ satisfies the crossing symmetry relation:
\ben
R_{12}(u)=M_2R^{t_1}_{12}(u^{-1} a )M_2 \quad \mbox{with} \quad M^2=\II  \quad \mbox{and} \quad
M_1M_2R_{12}(u)=R_{21}(u)M_1M_2. \label{CSG}
\een
then two families of reflection algebra are equivalent and 
\ben
\cK^*(u)=\cK(u \sqrt{a})M. \label{RAtoTRA}
\een

\vspace{0.2cm}
\subsection{The generalized $q$-Onsager algebra}
To solve the intertwiner equations (\ref{intRA}) or (\ref{intTRA}) one has to identify 
the $q$-Serre-Chevalley type formulation of the reflection algebras $\cB(\widehat{g})$ or $\cB^*(\widehat{g})$ respectively.

The generators made of $\cU_q(\widehat{g})$:
\ben
a_i=c_ie_i q_i^{h_i/2}+\bar{c}_if_i q_i^{h_i/2}+w_i q_i^{h_i}, \;\; \{c_i,\bar{c}_i,w_i\} \in \CC \label{genOq}\mb{and} i=0,1,\dots,rank(g)
\een
correspond to non-local conserved charges of the ATFT with non-preserving boundary conditions (parametrized by $w_i$) where $c_i,\bar{c}_i$ depend of the coupling constant of the theory \cite{MN,DM}. These charges close on the generalized $q$-Onsager algebras  $\cO_q(\widehat{g})$  upon certain restrictions on $w_i$, see bellow and \cite{BB}. 
These algebras are generated by $\{A_i\}$ with $i=0,1,\dots,rank(g)$ (see \cite{BB} for commutation relations) and have coideal properties given by the coaction:
\ben
\delta(A_i)=(c_ie_i q_i^{h_i/2}+\bar{c}_if_i q_i^{h_i/2})\otimes 1+ q_i^{h_i} \otimes A_i.
\een
and the counit:
\ben
\epsilon(A_i)=w_i.
\een
These coideal properties are necessary conditions for a $q$-Serre-Chevalley type formulation of  the reflection algebras $\cB(\widehat{g})$ or $\cB^*(\widehat{g})$ and appear in the definition of the intertwiner equations (\ref{intRA},\ref{intTRA}).
Moreover, the coideal properties $\delta,\epsilon$ fix uniquely the homomorphism $\psi: \cO_q(\widehat{g}) \to \cU_q(\widehat{g})$ from the relation $(1 \otimes \epsilon ) \circ \delta = \psi$ which implies that $\psi(A_i)=a_i$.

%
%

In the case of $\widehat{g}=a^{(1)}_1$ or $\widehat{g}=d^{(1)}_n$ these generators (\ref{genOq}) give, from intertwiner equation, the scalar solutions for the RA (\ref{intRA}) and in the case  $\widehat{g}=a^{(1)}_n, n>1,$ the scalar solutions for the TRA (\ref{intTRA}) \cite{DM,DelG}.
For the two first cases, the $R$-matrices satisfy the crossing symmetry (\ref{CSG}) and the solution of (\ref{intRA}) can be mapped on solution of (\ref{intTRA}) using (\ref{RAtoTRA}).
For the last case, the intertwiner equation (\ref{intRA}) with generators (\ref{genOq}) does not have non-trivial solutions.
In view of these results we can assume that the generalized $q$-Onsager algebras $\cO_q(\widehat{g})$ correspond to a $q$-Serre-Chevalley type formulation of the Twisted reflection algebras. 
For the case of $a^{(1)}_1$ the dynamical $K$-matrices have been obtained in terms of the $q$-Onsager algebra $\cO_q(a^{(1)}_1)$ \cite{BasS,BB2} and in terms of the Zhedanov or Askey-Wilson algebra\footnote{{The Zhedanov or Askey-Wilson algebra gives a realisation by a finite algebra of $\cO_q(a^{(1)}_1)$.}} 
$\cA\cW$ \cite{BK}. 
%

\vspace{0.2cm}

In this paper, we consider dynamical $K$-matrices with entries in a finite algebra that gives a realisation of $\cO_q(\widehat{g})$. The procedure to obtain these dynamical $K$-matrices, solutions of the reflection equation, is the following:

\vspace{0.1cm}
\begin{itemize}
\item[1)] Identify from the definition of the $q$-Onsager algebra $\cO_q(\widehat{g})$ its realisation by a finite algebra\footnote{ In most cases the finite algebra is given by a finite generalized $q$-Onsager algebra $\cO_q(g)$ isomorphic to a coideal subalgebra of $\cU_q(g)$ \cite{Le}, see also {\bf Remark 1}.}. In the cases considered in this paper the corresponding finite algebra is given by special cases of the  Zhedanov or Askey-Wilson algebra \cite{Zhed}.
\vspace{0.1cm}
\item[2)] Construct an irreducible basis of monomials of this algebra and define a 'minimal' solution of the $K$-matrix in terms of elements of the irreducible basis. The irreducibility of the basis ensures  that the solution is unique up to a scalar factor.
\vspace{0.1cm}
\item[3)] Solve the intertwiner equation (\ref{intTRA}) or (\ref{intRA}) using the commutation relations of the finite algebra.
\end{itemize}


\section{Dynamical $K$-matrices for  $\cO_{q}(a^{(2)}_{2})$ and  $\cO_{q}(a^{(1)}_{2})$}

In the present section we give dynamical (non-c-number or operator valued) $K$-matrices solutions of the intertwiner equations (\ref{intRA},\ref{intTRA}). We consider the cases corresponding to the generalised $q$-Onsager algebras $\cO_{q}(a^{(2)}_{2})$ and $\cO_{q}(a^{(1)}_{2})$.
At the first step, we introduce the Zhedanov or Askey-Wilson algebra and give a general ansatz for the $K$-matrix in terms of generators of this algebra. 
At the second step, we consider case by case the generalised $q$-Onsager algebras $\cO_{q}(a^{(2)}_{2})$ and $\cO_{q}(a^{(1)}_{2})$. We recall their definitions (infinite algebra) \cite{BB}, we give an evaluation homomorphism to the Zhedanov algebra (finite algebra) and, finally, we present the solution of the intertwiner equation - dynamical $K$-matrix. 

\vspace{0.2cm}

The Zhedanov algebra or Askey-Wilson algebra $\cA\cW(a,b,c,d,\bar{c},\bar{d},q)$ is given by:

\begin{defn}\cite{Zhed}
The Zhedanov algebra or Askey-Wilson algebra $\cA\cW(a,b,c,d,\bar{c},\bar{d},q)$ is an associative algebra generated by the elements $\{W_0,W_1,W_2\}$ and scalars $\{a,b,c,d,\bar{c},\bar{d},q\} \in \CC$ with $q$ (not root of unity), subject to the relations\footnote{We defined $[A,B]_q=AB-q^{-1}BA$.}:
\ben
&&[W_0,W_1]_{q}=a\,W_2,\\
&&\null[W_1,W_2]_{q}=b\, W_1+ c\,W_0+d, \\
&&\null[W_2,W_0]_{q}=b\, W_0+ \bar{c}\, W_1+\bar{d},\\
\een
The algebra is spanned by the monomials :
\ben
W_0^{i_0}W_1^{i_1}W_2^{i_2} \label{monob2}
\een
and has the Casimir :
\ben
Q &=& c q^{2}\,W_0^2+\bar{c} q^{-2}\,W_1^2+ a q^2 \,W_2^2+b(W_0W_1+W_1W_0) \nonumber \\
&& -q (q^{2}-q^{-2}) W_0\,W_1\,W_2+d q (q+q^{-1})\,W_0+\bar{d} q^{-1} (q+q^{-1}) W_1 \label{casb2}
 \een
\end{defn}

\begin{rmk} \label{rm-fOq}
For the generalized $q$-Onsager algebra $\cO_{q}(\widehat{g})$  a realization in terms of coideal subalgebra of the quantum algebra $\cU_q(g)$, with $g$ a simple Lie algebra, can be found. These coideal subalgebras  have been classified by
Letzter \cite{Le}. Among them one finds the Zhedanov algebra \cite{Zhed} and the Klimyk-Gavrilik algebra \cite{Klim,Gavr}. 
\end{rmk}

We look for a solution of the intertwiner equation (\ref{intRA},\ref{intTRA}) of the form:
\ben
K(u)=\sum_{i,j} E_{ij} \sum_{i_1,i_2,i_3} k^{i_1i_2 i_3} _{ij}(a,b,c,d,\bar{c},\bar{d},q,Q)W_0^{i_1}W_1^{i_2}W_2^{i_3}  \in End(\CC^3)\otimes \cW \label{ansatz}
\een
with positive integers $i_1+i_2+i_3 < p$, $ k^{i_1i_2 i_3} _{ij}(a,b,c,d,\bar{c},\bar{d},q,Q)$ some unknown functions and $\cW$ the vector space spanned by the monomials (\ref{monob2}).  
We choose $p=4$ similarly to the order of the Casimir  (\ref{casb2}). The irreducibility of the monomials\footnote{Proof of the irreducibility of the monomials can be found in \cite{Ter}.} (\ref{monob2}) ensures that the solution $K(u)$ is unique in each case and satisfies the corresponding reflection equation.

\subsection{The $\cO_{q}(a^{(2)}_{2})$ case: }

The affine Lie algebra $a^{(2)}_{2}$ is the simplest example of the twisted Kac-Moody algebras. We consider this example for its simplicity and its connection with the quantum integrable models as Bullough-Dodd, Mikhailov-Zhiber-Shabat or Izergin-Korepin models \cite{S,IK}. In this case there exists the crossing symmetry relation for the $R$-matrix of $\cU_q(a^{(2)}_{2})$ of the form (\ref{csA22}). Therefore, we can consider the intertwiner equation (\ref{intRA}) to construct solutions of the reflection equation (\ref{RA}). The solution for twisted reflection equation follows from the map (\ref{RAtoTRA}). 

The generalized $q$-Onsager algebra $\cO_{q}(a^{(2)}_{2})$ is defined by :

\begin{defn} \cite{BB} The generalized $q$-Onsager algebra $\cO_{q}(a^{(2)}_{2})$ is an associative algebra with unit 1, generated by the elements $\{A_{i}\}$, $i \in 0,1$ and scalars $\{\rho,\bar{\rho},\tilde{\rho}\} \in \CC$ subject to the relations:
\ben
[A_0,[A_0,A_1]_{q^{4}}]_{q^{-4}}-\rho A_1=0,\\
\null[A_1,[A_1,[A_1,[A_1,[A_1,A_0]_{q^{4}}]_{q^{-4}}]_{q^{2}}]_{q^{-2}}-\bar{\rho}\,(A_1\,A_1\,A_0-w\, A_1\,A_0\,A_1+A_0\,A_1\,A_1)-\tilde{\rho}\,A_0 ]=0,\\
w=\frac{(q-1+q^{-1})(q^{4}+2\,q^{2}+4+2\,q^{-2}+q^{-4})}{q^{4}+3+q^{-4}}.
\een
It is a coideal subalgebra with counit $\epsilon : \cO_{q}(a^{(2)}_{2}) \to \CC$ given by :
\ben
\epsilon(A_0)= w_0, 	\quad \epsilon(A_1)= w_1\quad  \mbox{with} \quad \Big(w_0^2+\frac{\rho}{(q^2-q^{-2})^2}\Big)w_1=0. \label{contw1}
\een
The coaction $ \delta : \cO_{q}(a^{(2)}_{2}) \to \cU_{q}(a^{(2)}_{2})\otimes \cO_{q}(a^{(2)}_{2})$ is given by :
\ben
\delta(A_i)=  (c_i\,e_iq_i^{\frac{h_i}{2}} +\overline{c}_i\,f_iq_i^{\frac{h_i}{2}})\otimes \II + q_i^{h_i} \otimes A_i, \label{coactionmap}
\een
with 
\ben
\rho=c_0 \overline{c}_0, \quad \bar{\rho}=(q+q^{-1})^2(q^4+3+q^{-4})c_1 \overline{c}_1
\quad \mbox{and} \quad \tilde\rho=-\left(\frac{(q^2+q^{-2}) \bar\rho}{(q^4+3+q^{-4})}\right)^2,
\een
\end{defn}

and we have the evaluation homomorphism $\Pi$:

 \begin{prop}
There is an algebra homomorphism $\Pi : \cO_q(a^{(2)}_{2})\ \rightarrow \cA\cW(1,0,-q^{-4} \bar{\mu},C,-q^{-4}{\mu},0,q^4)$ such that 
\ben
\Pi(A_i)=W_i \mb{for} i\in\{0,1\} \mb{and} \Pi(\rho)=\mu, \quad \Pi(\bar\rho)=\frac{(q^4+3+q^{-4}) \bar{\mu}}{(q^2+q^{-2})^2 }. \label{realaw1}
\een
\end{prop}

\vspace{0.2cm}

For convenience let us use the following new parameters for $\cA\cW$ and the matrix representation of $\cU_q(a^{(2)}_{2})$:
\ben
\alpha =\sqrt{\mu}, \quad 
\beta=\frac{q^4}{\sqrt{\bar{\mu}(q+q^{-1})}}, \quad
s_0=q^2 \sqrt{\frac{\bar{c}_0}{c_0}}, \quad s_1= \sqrt{\frac{q(q+q^{-1})\bar{c}_1}{c_1}} \mb{and} c_i,\bar{c}_i \in \RR^+ 
\een 
From straighforward calculations of the intertwiner equation (\ref{intRA}) we obtain the dynamical $K$-matrix: 
\ben
K(u)= K^{(1)}(u)+ K^{(2)}(u) \label{KA22dyn}
\een
with
\ben
&&K^{(1)}(u)=\begin{pmatrix}
h_{1}(u,C,Q) &0 & h(u,C)\\
0 &h_{2}(u,C,Q)& 0\\
h(u,C)&0 &h_{3}(u,C,Q)
\end{pmatrix}\nonumber,\\
&&{\footnotesize K^{(2)}(u)=\begin{pmatrix}
\alpha \beta g\,u\, W_1^2+\frac{q}{\beta(q+q^{-1})}W_0 &q^{-\frac{7}{2}}[W_1,W_0]_{q^{-4}}+q^{\frac{7}{2}}\alpha\, u\, W_1 & \beta \, \null[\null[W_0,W_1]_{q^{4}},W_1]_{q^{4}}\\
q^{-\frac{7}{2}}\null[W_0,W_1]_{q^{-4}}+q^{\frac{7}{2}}\alpha\,  u \, W_1 &-\frac{q^{4}}{\beta}W_0& q^{\frac{7}{2}}\null[W_0,W_1]_{q^{4}}+q^{-\frac{7}{2}}\alpha\,  u^{-1}\,  W_1\\
\beta \, \null[W_1,\null[W_1,W_0]_{q^{4}}]_{q^{4}}&q^{\frac{7}{2}}\null[W_1,W_0]_{q^4}+q^{-\frac{7}{2}}\alpha\,  u^{-1}\, W_1&-\alpha \beta \frac{g}{q^8\,u}\,W_1^2+\frac{q^{7}}{\beta(q+q^{-1})}W_0
\end{pmatrix}}\nonumber\een
and 
\ben
h(u,C)&=&\frac{\alpha q^{4} (q^3 u-q^{-3} u^{-1})}{\beta  (q-q^{-1}) (q^4-q^{-4})}-\beta C \nonu
h_{1}(u,C,Q)&=& \frac{\beta Q (q^4-q^{-4})}{\alpha (u-u^{-1})}+ \frac{ u \beta C (q-q^{-1})(q^2+q^{-2})}{(u-u^{-1})} +\frac{ q^4 u \alpha (q^3 u-q^{-3} u^{-1})}{\beta (u-u^{-1})(q+q^{-1})^2(q^2+q^{-2})} \nonu
h_{2}(u,C,Q)&=& \frac{\beta Q (q^4-q^{-4})}{\alpha (u-u^{-1})}+ \frac{\beta C (u\,q-u^{-1}\,q^{-1})(q^2+q^{-2})}{ (u-u^{-1})} +\frac{ q^4  \alpha (q u-q^{-1} u^{-1}) (q^3 u-q^{-3} u^{-1})}{\beta (u-u^{-1})(q+q^{-1})(q^4-q^{-4})}\nonu
h_{3}(u,C,Q)&=& \frac{\beta Q (q^4-q^{-4})}{\alpha (u-u^{-1})}+ \frac{\beta C (q-q^{-1})(q^2+q^{-2})}{u (u-u^{-1})} +\frac{ q^4  \alpha (q^3 u-q^{-3} u^{-1})}{u \beta (u-u^{-1})(q+q^{-1})^2(q^2+q^{-2})} \nonu
g&=&(q^4-q^{-4}).\nonumber
\een

The direct consequences of the intertwiner equation (\ref{intRA}) are that the inverse of $K(u)$ is proportional to $K(u^{-1})$, the $K$-matrix is unique (up to a scalar function) and is a solution of the reflection equation:
\ben
R'_{12}\Big(\frac{u}{v}\Big) K_1(u) R'_{21}(u v)  K_2(v)=  K_2(v)R'_{12}(u v) K_1(u) R'_{21}\Big(\frac{u}{v}\Big) \label{REA22}
\een
with
\ben
R'_{12}(u)=R_{21}(u,q).
\een
The definition and main properties of the $R$-matrix are given in the appendix A.

\vspace{0.2cm}


\subsection{The $\cO_{q}(a^{(1)}_{2})$ case}

The affine Lie algebra $a^{(1)}_{2}$ is the simplest example of Kac-Moody algebra of rank three and as in the previous case is connected with quantum affine Toda fields theories \cite{Ga}. The $R$-matrix of its deformation $\cU_q(a^{(1)}_{2})$ does not have the  
crossing symmetry (\ref{CSG}) and it means that the two families of the reflection equations are different. Moreover, the evaluation of $\cK^*(u)$\footnote{The symbol $*$ here indicates that the $K$-matrix is a solution of the twisted reflection equation (\ref{TRA}).} in a finite dimensional algebra is known for the case $a^{(1)}_{n}$ \cite{MRS} and supports the fact that  $\cO_{q}(a^{(1)}_{2})$  is a q-Serre-Chevalley formulation of TRA.

\vspace{0.2cm}

The generalised $q$-Onsager algebra $\cO_{q}(a^{(1)}_{2})$ is given by the definition:

\begin{defn} \cite{BB} The generalized $q$-Onsager algebra $\cO_{q}(a^{(1)}_{2})$ is an associative algebra with unit 1, generated by the elements $\{\sA_{i}\}$ and scalars $\rho_i \in \CC$ with $i=0,1,2$ subject to the relations:
\ben
\null[A_i,[A_i,A_{i\pm1}]_{q}]_{q^{-1}} &=& \rho_i A_{i\pm1}, \mb{for} i=0,1,2 \quad(3 \equiv 0, \, -1 \equiv 2) \label{Oqa12} 
\een
It is a coideal subalgebra with counit, $\epsilon : \cO_{q}(a^{(1)}_{2}) \to {\mathbb C}$ given by:
\ben
\epsilon(A_i)= w_i, \mb{with} \Big(w_i^2+\frac{\rho_i}{q+q^{-1} - 2}\Big)w_j = 0 \mb{for any} i,j \in \{0,1,2\} \label{contw2}
\een
and the coaction $\delta : \cO_{q}(a^{(1)}_{2}) \to \cU_{q}(a^{(1)}_{2})\otimes \cO_{q}(a^{(1)}_{2})$ is given by:
\ben
\delta(A_i)=  (c_i\,e_i q^{\frac{h_i}{2}} +\overline{c}_i\,f_i q^{\frac{h_i}{2}})\otimes I\!\!I + q^{h_i} \otimes A_i \label{coactionmap}
\een
with
 \ben
 \{c_i, \overline c_i\} \in {\CC} \mb{and}   \rho_i = c_i \overline c_i 
  \een
\end{defn}

and we have the evaluation homomorphism $\Pi$:

\begin{prop}
There is an algebra homomorphism $\Pi : \cO_q(a^{(1)}_2)\ \rightarrow \cA\cW(\mu_0,0,-\frac{\mu_2}{q \mu_0},0,-\frac{\mu_1}{q \mu_0},0,q)$ such that 
\ben
\Pi(A_j) = W_{j}    \mb{for} j \in\{0,1,2\}, \quad
\Pi(\rho_1) =  \mu_1, \quad
\Pi(\rho_2) =  \mu_2 \mb{and} 
\Pi(\rho_0) =  q^{-1} \frac{\mu_1 \mu_2}{( i \mu_0)^2}, \;\; \label{realaw2} 
\een
\end{prop}

\vspace{0.2cm}

In this case, we consider the intertwiner equations (\ref{intTRA}) for the $K$-matrices $K^*(u)$ transforming the representation $\pi$ into $\bar{\pi}$ and also the interwiner :
\ben
\bar K^*(u)(\bar \pi_u \otimes \Pi ) \delta(y)=({\pi}_{u^{-1}} \otimes \Pi) \delta(y) \bar K^*(u) , \quad y \in B^*(\widehat{g})  \label{intTRAI}
\een
transforming the representation $\bar{\pi}$ into $\pi$. The two interwiners are related by the fact that   $\bar K^*(u^{-1})$ is proportional to the inverse of $K^*(u)$.
For convenience let us use the following new parameters for $\cA\cW$ and the matrix representation of $\cU_q(a^{(1)}_{2})$:
\ben
&& s_i =  \sqrt{q \overline c_i / c_i  } , \quad \alpha_0= \sqrt{\mu_0}, \quad \alpha_1= i \sqrt{\frac{\mu_1}{q \mu_0}} , \quad \alpha_2= i \sqrt{\frac{\mu_2}{q \mu_0}}, \quad  \alpha=  \alpha_1 \alpha_2 \alpha_3  \mb{and} c_i,\bar{c}_i \in \RR^+ 
\een

The solution of the first intertwiner equation, (\ref{intTRA}), is given by  
\ben
\quad \qquad K^*(u) = 
\left(
\begin{array}{ccc}
		\alpha \kappa(u)  & 
		-q^{-\frac{1}{2}} u^{-1} \alpha_1 W_1 &
		i u^{-1}  \alpha_0 W_0 
		\\
		i  u \, \alpha_1 W_1 & 
		\alpha \kappa(u)  &
		-q^{-\frac{1}{2}} u^{-1} \alpha_2 W_2 
		\\ 
		-q^{-\frac{1}{2}} u \, \alpha_0 W_0 &
		i u \, \alpha_2 W_2 &  
		\alpha \kappa(u) 
\end{array}
\right) \label{KA12dyn}
\een
where $\kappa(u) =\frac{q^{\frac{1}{2}} u - i u^{-1} }{q-q^{-1}} $. 
We remark that there is another solution given by the transformation $i \to -i$ leaving the homomorphism of {\bf Proposition 3.2} invariant. 

\vspace{0.2cm}

The solution for the second intertwiner equation (\ref{intTRAI}), $\bar K^*(u)$, is given by:
\ben
\bar K^*(u) = \bar K^*_{(1)}(u) +\frac{i}{q-q^{-1}} \bar K^*_{(2)}(u), \label{KA12dyn2}
\een
with
\ben
&& \bar K^*_{(1)}(u) =  \left(
\begin{array}{ccc}  \alpha \, \bar{\kappa}(u)^2- \frac{i q^{\frac{1}{2}}}{\alpha}\, (\alpha_2 W_2)^2 & 0 & 0 \\
0 &  \alpha \, \bar{\kappa}(u)^2- \frac{i q^{\frac{1}{2}}}{\alpha}\, (\alpha_0 W_0)^2 & 0 \\
0 & 0 & \, \alpha \, \bar{\kappa}(u)^2- \frac{i q^{\frac{1}{2}}}{\alpha}\,(\alpha_1 W_1)^2\end{array}
\right)
 \nonumber 
\\
&& \bar K^*_{(2)}(u) = \left(
\begin{array}{ccc}
	0 &
   q^{\frac{5}{2}} u^2 \alpha_1 W_1- \frac{i}{\alpha_1} [W_0,W_2]_q & 
  q^{-1/2} \alpha_0 W_0 -  \frac{i q u^2}{\alpha_0} [W_1, W_2]_{q^{-1}} 
\\
	q^{-\frac{1}{2}} u^{-2} \alpha_1 W_1- \frac{ i q}{\alpha_1} [W_2,W_0]_{q^{-1}} & 
0   &
	q^{\frac{5}{2}} u^2 \alpha_2 W_2- \frac{i q}{\alpha_2}[W_1,W_0]_{q}  
\\
	q^{\frac{5}{2}} \alpha_0 W_0- \frac{i u^{-2}}{\alpha_0} [W_2, W_1]_{q} &
  q^{-\frac{1}{2}} u^{-2} \alpha_2 W_2- \frac{i q}{\alpha_2} [W_0,W_1]_{q^{-1}}  &
0
\end{array}
\right) \nonumber
\een
and $\bar{\kappa}(u) =\frac{q^{\frac{3}{2}} u + i u^{-1} }{q-q^{-1}} $. Similarly, there is another solution which is also given by the transformation $i \to -i$. 

\vspace{0.2cm}

Finally, these solutions, $K^*(u)$ and $\bar K^*(u)$, are unique (up to a scalar function) and satisfy the twisted reflection equations (\ref{TRA}) \footnote{The second twisted reflection equation corresponds to the inverse of the first one (up to a scalar function) and  $K^*(u)\bar K^*(1/u)=\bar K^*(1/u)K^*(u) \propto 1$.}: 
\ben
R_{12} (u/v) \cK^*_1(u) R^{t_1}_{12} (\frac{1}{u v}) \cK^*_2(v)  =  \cK^*_2(v) R^{t_1}_{12} (\frac{1}{u v}) \cK^*_1(u) R_{12} (u/v) \\
R_{21} (u/v) \bar \cK^*_1(u) R^{t_2}_{12} (\frac{1}{ u v}) \bar \cK^*_2(v) =  \bar \cK^*_2(v) R^{t_2}_{12} (\frac{1}{u v}) \bar \cK^*_1(u) R_{21} (u/v) 
\een
with $\cK^*(u) = V K^*\big( u \sqrt{i} q^{-\frac{3}{4}}\big)$ and $\bar \cK^*(u) = \bar K^*\big(u  \sqrt{i} q^{-\frac{3}{4}}\big) V$. The $R$-matrix and the matrix $V$ are defined in the appendix A.
 
\begin{rmk}
Using the invariance of the twisted reflection equation by the map $\cK^*(u) \to M \cK^*(u) M$, with $M$ an arbitrary diagonal matrix, and rescaling the generators $\{W_i\}$  we found that our solution (\ref{KA12dyn}) coincides with the result obtained in \cite{MRS} for the orthogonal case.
\end{rmk}


\section{Scalar limit of the dynamical $K$-matrices}

Starting from our dynamical solutions (\ref{KA22dyn},\ref{KA12dyn},\ref{KA12dyn2}), it is straighforward to derive the already known scalar solutions of the
reflection equation (\ref{RA}) and twisted reflection equations (\ref{TRA}).
These solutions have been derived directly from the reflection equation \cite{HB,K,BFKZ} or from matrix intertwiner operator \cite{N}.
We obtain the scalar solutions (c-number $K$-matrices) from the dynamical solutions (non-c-number $K$-matrices) taking the trivial, or one dimensional representation of $\cA\cW$.
\begin{rmk} 
The scalar solutions obtained from the dynamical ones are the same than the ones obtained from the intertwiner equations (\ref{intRA},\ref{intTRA}) replacing $\Pi$ by $\epsilon$.
\end{rmk}
 
For the case $a^{(2)}_2$, the scalar representation of  $W_0$ and $W_1$ must be solutions of the equation (\ref{contw1}). It follows two types of scalar $K$-matrices:

\begin{itemize}
\item The case $w_1=0$,  $w_0$ arbitrary and 
\ben
w_2=0, \quad C=\frac{q^{4}w_0}{\beta^2(q+q^{-1})}, \quad Q= -\frac{q^{4}w^2_0}{\beta^2(q+q^{-1})}\nonumber
\een
leads to the Type II scalar solution in \cite{K,N,L-S}

\ben
K^{scal}_{w_0,0}(u)&\propto&\II+\frac{\frac{\alpha}{w_0}}{(q+q^{-1})(q^{2}+q^{-2})}\begin{pmatrix} 
u & 0 & \frac{u-u^{-1}}{q-q^{-1}}   \\
0 &\frac{q\,u-q^{-1}\,u^{-1}}{q-q^{-1}}    & 0 \\
 \frac{u-u^{-1}}{q-q^{-1}}  & 0 & u^{-1} 
\end{pmatrix}\nonumber
\een
The additional boundary parameters in \cite{K,N} are related to the invariance of the reflection equation by conjugation by the diagonal matrix of the form $M=Diag(a_1,(a_1a_2)^{1/2},a_2)$. 
The trivial solution proportional to unit can be recovered from this solution taking $\alpha=0$.
\\
\item The cases $w_0=\pm i \frac{\alpha}{q^2-q^{-2}}$, $w_1$ arbitrary,
\ben
&&w_2=\pm i \alpha w_1q^{-2}, \quad C=\pm i \alpha\left( w_1^2q^{-4}(q^2-q^{-2})+q^4\frac{1}{\beta^2(q+q^{-1})(q^2-q^{-2})}\right),\nonu
&&Q= \alpha^2 \left((q^4+q^{-4}) w_1^2+ \frac{q^4}{\beta^2 (q+q^{-1})(q^2-q^{-2})^2}\right)\nonumber
\een
lead to the scalar solutions with all non-zero entries or Type I in \cite{N}
\ben
&&K^{scal}_{\pm i \frac{\alpha}{q^2-q^{-2}},w_1}(u)\propto K^{scal}_{(1)}(u)+\frac{q^7}{(\beta w_1)^2 [4]_q^2(q-q^{-1})(q^3\,u\pm i) }K^{scal}_{(2)}(u)\nonumber\\
&&K^{scal}_{(1)}(u)=\begin{pmatrix} 
u & 0 & 0  \\
0 &\frac{q^{3}\pm i\,u}{q^3\,u\pm i}    & 0 \\
 0 &0  & u^{-1} 
\end{pmatrix}\nonumber\\
&&K^{scal}_{(2)}=\begin{pmatrix} 
\frac{q^2(q\,u\mp i)((q-q^{-1})u\pm(q^2+q^{-2}))}{(q^2-q^{-2})} &\beta w_1 u q^{-\frac{1}{2}}(u-u^{-1})[4]_q  & \frac{q^2(u-u^{-1})(q \,u\mp i)}{(q^2-q^{-2})}    \\
\beta w_1 u q^{-\frac{1}{2}} (u-u^{-1})[4]_q   & \frac{(q\,u\mp i)(u\pm i\,q)(q^3 \,u\pm i)}{u(q^2-q^{-2})}   &\pm i\,\beta w_1\,q^{-\frac{5}{2}}  (u-u^{-1})[4]_q   \\
\frac{q^2(u-u^{-1})(q \,u\mp i)}{(q^2-q^{-2})}    & \pm i\,\beta  w_1\,q^{-\frac{5}{2}} (u-u^{-1})[4]_q  & \frac{q^2(1\pm i\,q\,u)((q^2+q^{-2})u\mp(q-q^{-1})i)}{u (q^2-q^{-2})} 
\end{pmatrix}\nonumber
\een
The two others non-trivial diagonal solutions can be recovered from these solutions taking $\beta \to \infty$.
 \end{itemize}
Let us remark that the number of free parameters in the scalar solution is in one-to-one correspondence with the trivial representations of $\cO_q(a^{(2)}_2)$, all the other parameters appearing in the previous solution correspond to matrix conjugation and multiplication by a scalar that leaves the reflection equation invariant. 

 \vspace{0.2cm}
 
For the case $a^{(1)}_2$, the scalars $w_i$ are more restricted by $ \cA\cW(\mu_0,0,-\frac{\mu_2}{q \mu_0},0,-\frac{\mu_1}{q \mu_0},0,q)$ than by $\cO_q(a^{(1)}_2)$. But the invariance of the homomorphism of {\bf Proposition 3.2} by the transformation $i\to -i$ leads to the same number of scalar solutions as for $\cO_q(a^{(1)}_2)$ trivial representations:

\begin{itemize}	
\item
Let $e_i=\pm 1$ for $i=1,2$ or $e_1=e_2=0$, then scalar representation of $\cO_q(a_2)$ are given by:
\ben
w_0=e_1e_2 q \frac{ \sqrt{\mu_1 \mu_2}}{q-1}, \quad w_0=e_1 q \frac{ \sqrt{\mu_0 \mu_2}}{q-1} \mb {and} w_2=e_2 q \frac{ \sqrt{\mu_0 \mu_1}}{q-1}.
\een
 Considering the both solutions, i.e $i\to e_0 i$ with $e_0=\pm1$, we recover the scalar solutions \cite{Ga,DM}:
\ben
K^{scal}_{e_0,e_1,e_2}(u)\propto \begin{pmatrix}\frac{q^{\frac{1}{2}} u -  i e_0 u^{-1} }{q^{1/2}+q^{-1/2}} & -  u^{-1} e_1 &  i q^{\frac{1}{2}} u^{-1} e_0 e_1 e_2 \\
 i q^{\frac{1}{2}} u e_0e_1 & \frac{q^{\frac{1}{2}} u - i e_0 u^{-1} }{q^{1/2}+q^{-1/2}}  &  -  u^{-1} e_2 \\
 -  u e_1 e_2 & i q^{\frac{1}{2}} u e_0 e_2 &  \frac{q^{\frac{1}{2}} u -  i e_0 u^{-1} }{q^{1/2}+q^{-1/2}}   \end{pmatrix}
 \een
 \end{itemize}

\section{Comments}

The dynamical $K$-matrices solutions of the RA and the TRA have been constructed for the case $a_{2}^{(2)}$ and $a_2^{(1)}$ respectively using the generalized $q$-Onsager algebras and the intertwiner equations. These $K$-matrices have entries in the finite subalgebra of the related generalized $q$-Onsager algebra (some special cases of Zhedanov algebra \cite{Zhed}) and are solutions of the reflection equations. Taking the trivial representation of these subalgebras we recover the known scalar solutions. 

\vspace{0.2cm}

This procedure can clearly be applied to other generalized $q$-Onsager algebras related to $\widehat{g}$ and corresponding dynamical $K$-matrices can be obtained. The main problem is to identify a realisation of the generalised q-Onsager $\cO_q(\widehat{g})$ in terms of a finite algebra and construct an irreducible basis for the vector space spanned by monomials of generators of the algebra.
As we already mentioned, the case of $\cO_q(a^{(1)}_{n-1})$ is already known starting directly from TRA \cite{MRS} and the corresponding finite algebra is the Klimyk-Gavrilik algebra \cite{Klim,Gavr}. 
 It corresponds to the orthogonal case of the TRA (noted $Y^{tw}_q(o_n)$ in  \cite{MRS}).
This fact suggests that $\cO_q(a^{(1)}_{n-1})$ corresponds to a $q$-Serre-Chevalley formulation of $Y^{tw}_q(so_n)$ (quotient of $Y^{tw}_q(o_n)$ by its center).
For the case $Y^{tw}_q(sp_n)$, also considered in \cite{MRS}, the $q$-Serre-Chevalley formulation still remains, to our knowledge, as an open problem and will be considered elsewhere.

\vspace{0.2cm}

In the case of a finite Lie algebra $g$, the associated reflection algebras have been studied in the context of the quantization of the symmetric spaces and related q-orthogonal polynomial \cite{NS,Le}. 
Two remarkable facts for us appear in these works: there is a one-to-one correspondence between scalar solutions of the reflection equations and the quantum symmetric spaces \cite{NS};  the classification of the coideal subalgebras of $\cU_q(g)$ is obtained by the classification of the irreducibles pair $(g,\theta)$, with $\theta$ an involution of $g$, and the $q$-Serre-Chevalley formulation follows from the involution and the coideal properties \cite{Le}. The extension of this classification for the affine case is still an open problem.

\vspace{0.2cm}

As for physical applications in quantum affine Toda fields theories with boundary degree of freedom, 
the dynamical solutions obtained in this paper give the algebraic part of the boundary scattering amplitudes for fundamental particles in the $a_{2}^{(2)}$ and $a_2^{(1)}$ theories with imaginary coupling. Next step is to consider the fusion procedure and obtain the boundary scattering amplitudes for fused particles. As an application, they lead to scattering amplitudes of the fundamental particles of the real
coupling theory. In the case of the $a^{(1)}_1$ boundary affine Toda field theory with real coupling (called sinh-Gordon model)
and dynamical boundary conditions \cite{BK}, a remarkable weak-strong coupling duality property of boundary amplitudes is observed.
By analogy, weak-strong coupling duality properties could be investigated for $a_{2}^{(2)}$ and $a_2^{(1)}$ cases using the results presented here.
Moreover, considering infinite dimensional representation of $\cA\cW$ algebra, it can be interesting to compare our dynamical solutions with the ones obtained from scalar solutions dressed by equations (\ref{dressRA},\ref{dressTRA}) with bulk amplitudes of impurities \cite{CZ}. These questions will be considered elsewhere.

\vspace{1 cm}

\noindent{\bf Acknowledgements:} The authors thank P.Baseilhac for the discussions and valuable comments, the LMPT of Tours for hospitality where part of this work has been done. V.F. thanks INFN of Bologna for hospitality where part of this work has been done. S.B. was supported by  INFN Iniziativa Specifica FI11. 

\vspace{1 cm}

\begin{center} {\bf Appendix A : $U_{q}(a^{(2)}_{2})$ and $U_{q}(a^{(1)}_{2})$  quantum algebras}\end{center}

\vspace{0.5cm}

In this appendix we give the definitions of the $U_{q}(a^{(2)}_{2})$ and $U_{q}(a^{(1)}_{2})$ algebras, their fundamental representations and the associated $R$-matrices used in this paper.  
The generalised Cartan matrix $A=(a_{ij})_{i,j=0,n}$ and associated coprime positive integers $d_0,..,d_n$ are given,  respectively, for $a^{(2)}_{2}$ and $a^{(1)}_{2}$ by:
\ben
A = 
\begin{pmatrix} 
2 & -1 \\
-4 & 2
\end{pmatrix},\quad
 (d_{i})=\{
4,1\}, \mb{and} A = 
\begin{pmatrix} 
2 & -1 & -1 \\
-1 & 2 & -1 \\
-1 & -1 & 2
\end{pmatrix}
,\quad
 (d_{i})=\{
1,1,1\}.\nonumber
\een
\\
{\bf Definition A.1.}  {\it The quantum algebra $U_q(\widehat{g})$ is an associative algebra over ${\mathbb C}$ with unit $1$ generated by the elements
$\{ e_i, f_i,q_i^{\pm \frac{h_i}{2}}\}$, $i \in 0,..,n$ subject to the relations:
\ben
q_i^{\pm\frac{h_i}{2}}q_i^{\mp \frac{h_i}{2}}=1, \quad
q_i^{\frac{h_i}{2}}q_j^{\frac{h_j}{2}}=q_j^{\frac{h_j}{2}}q_i^{\frac{h_i}{2}},\quad
q_i^{\frac{h_i}{2}}\,e_j\,q_i^{-\frac{h_i}{2}}=q_i^{\frac{a_{ij}}{2}}\,e_j, \quad
q_i^{\frac{h_i}{2}}\,f_j\,q_i^{-\frac{h_i}{2}}=q_i^{-\frac{a_{ij}}{2}}\,f_j ,\nonu
\null [e_i,f_j]=\delta_{ij}\frac{q_i^{h_i}-q_i^{-h_i}}{q_i-q_i^{-1}}\ ,\nonu
\sum_{r=0}^{1-a_{ij}}(-1)^r
\left[ \begin{array}{c}
1-a_{ij} \\
r 
\end{array}\right]_{q_i}
e_i^{1-a_{ij}-r}\,e_j\,e_i^r=0\ , \quad \sum_{r=0}^{1-a_{ij}}(-1)^r
\left[ \begin{array}{c}
1-a_{ij} \\
r 
\end{array}\right]_{q_i}
f_i^{1-a_{ij}-r}\,f_j\,\,f_i^r=0\ .\nonumber
\een

with $q_i=q^{d_i}$.
The Hopf algebra structure is ensured by the existence of a comultiplication $\Delta: U_{q}(\widehat{g})\mapsto U_{q}(\widehat{g})\otimes U_{q}(\widehat{g})$, antipode ${\cal S}: U_{q}(\widehat{g}\mapsto U_{q}(\widehat{g})$ and a counit ${\cal E}: U_{q}(\widehat{g})\mapsto {\mathbb C}$ :}
\ben \Delta(e_i)&=&e_i\otimes q_i^{-h_i/2} +
q_i^{h_i/2}\otimes e_i\ , \quad
\Delta(f_i)=f_i\otimes q_i^{-h_i/2} + q_i^{h_i/2}\otimes f_i, \quad
\Delta(h_i)=h_i\otimes I\!\!I + I\!\!I \otimes h_i\ ,\nonumber
\een
\beqa {\cal S}(e_i)=-q_i^{-1}e_i\ ,\quad {\cal S}(f_i)=-q_i f_i\ ,\quad {\cal S}(h_i)=-h_i \ ,\qquad {\cal S}({I\!\!I})=1,\
\label{antipode}\nonumber\eeqa
\beqa {\cal E}(e_i)={\cal E}(f_i)={\cal
E}(h_i)=0\ ,\qquad {\cal E}({I\!\!I})=1\
.\label{counit}\nonumber\eeqa

For both cases, we use three-dimensional representations, constructed with the unit matrices $E_{ij}$ with $1$ at the intersection of the line $i$ and the column $j$ and zero elsewhere. 

\vspace{0.1cm}

$\bullet$ For $U_{q}(a^{(2)}_{2})$ case the fundamental representation $\pi_{u} : U_{q}(a^{(2)}_{2})\mapsto \CC^3 [[u,u^{-1}]]$ is given by \cite{FRS}:
\beqa
&\pi_{u}(e_{1}) =s_1(E_{12} +E_{23}),\;&\; \pi_{u}(e_{0}) =u s_0 E_{31}, \nonu
&\pi_{u}(f_{1}) = (q+q^{-1})s^{-1}_1(E_{21} + E_{32}),\;&\; \pi_{u}(f_{0}) =u^{-1} s_0^{-1}E_{13}, \nonu
&\pi_{u}\big(q_1^{\frac{h_1}{2}}\big) =q E_{11} + E_{22} + q^{-1}E_{33},\;&\; \pi_{u}\big(q_0^{\frac{h_0}{2}}\big) = q^{-2} E_{11} + E_{22} + q^{2}E_{33}\nonumber.
\eeqa
The parameters $s_i$ correspond to an automorphism\footnote{ given by $e_i\to s_i e_i, f_i\to \frac{1}{s_i} f_i$ and $h_i \to h_i$.}  of $U_{q}(a^{(2)}_{2})$ irrelevant for the $R$-matrix but relevant for the $K$-matrix. 
Following the results of Jimbo \cite{J} we can derive, up to a scalar function, the $R$-matrix associated to this representation from the intertwining property:
 The solution is given by:
\ben
R_{12}(u)=(u-1)q^3 R_{21}^{{(q)}}+(1-u^{-1}) q^{-3} R_{12}^{(q^{-1})}+q^{-5}(q^4-1)(q^6+1)P  \label{RA22}
\een
where $R_{12}^{(q)}$ is the spin one or three-dimensional $R$-matrix associated to the $U_q(a_1)$ algebra given by :
\ben
R_{12}^{(q)}=\begin{pmatrix} 
q^{2} & 0  & 0 & 0  & 0 & 0  & 0 & 0  & 0 \\
0 & 1  & 0 & q^{2}-q^{-2}  & 0 & 0  & 0 & 0  & 0 \\
0 & 0  & q^{-2} & 0  &  q (q^{2}-q^{-2}) & 0  & q (q^{2}-q^{-2})(q-q^{-1}) & 0  & 0 \\
0 & 0  & 0 & 1  & 0 & 0  & 0 & 0  & 0 \\
0 & 0  & 0 & 0  & 1 & 0  & q (q^{2}-q^{-2}) & 0  & 0 \\
0 & 0  & 0 & 0  & 0 & 1 & 0 & q^{2}-q^{-2}  & 0 \\
0 & 0  & 0 & 0  & 0 & 0  & q^{-2} & 0  & 0 \\
0 & 0  & 0 & 0  & 0 & 0  & 0 & 1  & 0 \\
0 & 0  & 0 & 0  & 0 & 0  & 0 & 0  & q^{2} \\
\end{pmatrix}\nonumber
\een
It follows that this $R$-matrix is a solution of the Yang-Baxter equation (\ref{YB}).
It also satisfy the unitarity condition
\ben
R_{12}(u)R_{21}(u^{-1})=q^{2} (1-q^{4}u)(1+q^{-6}u)(1-q^{4}u^{-1})(1+q^{-6}u^{-1})\,\II 
\een
and the crossing symmetry:
\ben
R_{12}(u)=M_1R^{t_2}_{21}(-q^{-6}u^{-1}) M_1\label{csA22}
\mb{with}
M=-qE_{31}+E_{22}-q^{-1}E_{13}.
\een

\vspace{0.1cm}

$\bullet$ For $U_{q}(a^{(1)}_{2})$ case there are two fundamental representations $\pi_{u},\bar{\pi}_{u} : U_{q}(a^{(1)}_2)\mapsto \CC^3 [[u,u^{-1}]]$ given by:
 \ben
\pi_u (e_0) &=& u^2 s_0 E_{31}, \quad \pi_u (f_0) = u^{-2} s_0^{-1} E_{13}, \quad \pi_u (q^{h_0}) = q^{-1} E_{11} + E_{22} + q E_{33}, \\
\pi_u (e_1) &=& s_1 E_{12}, \quad \pi_u (f_1) = s_1^{-1} E_{21}, \quad \pi_u (q^{h_1}) = q E_{11} + q^{-1} E_{22} + E_{33}, \\
\pi_u (e_2) &=& s_2 E_{23}, \quad \pi_u (f_2) = s_2^{-1} E_{32}, \quad  \pi_u (q^{h_2}) = E_{11} + q E_{22} + q^{-1} E_{33}. 
\een
and 
\ben
\bar{\pi}_u (g) = (\pi_u(g))^t |_{q \rightarrow 1/q} \mb{for any} g \in U_q(a^{(1)}_2)
\een
where $t$ means the transposition. The $R$-matrix defined from intertwiner equation with two $\pi$ representations is given by
\ben
R_{12}(u) &=& \sum_{i} E_{ii} \otimes  E_{ii} + \frac{u-u^{-1}}{q u - q^{-1}u^{-1}} \sum_{{i \neq j}} E_{ii} \otimes E_{jj} + \frac{q - q^{-1}}{q u - q^{-1}u^{-1}}  \sum_{{i \neq j}} u^{sign(i-j)} E_{ij} \otimes E_{ji}. \label{RA12}
\een
This $R$-matrix is a solution of the Yang-Baxter equation (\ref{YB})
and also satisfy the unitarity condition
\ben
R_{12}(u) R_{21}(1/u) = \II, 
\een
The $R$-matrix defined from intertwiner equation with two $\bar{\pi}$ representations is given by $R_{12}^{t_1t_2}(u)=R_{21}(u)$ and is also a solution of (\ref{YB}).
The remaining cases $\pi_u \otimes \bar{\pi}_v$ and $\bar{\pi}_u \otimes \pi_v$ are given, respectively, by:
\ben
\qquad V^{-1}_{1} R^{t_{2}}_{21}( i q^{-3/2} u^{-1} )  V_{1} \mb{and}  V_1 R^{t_1}_{21}( i q^{-3/2} u^{-1} ) V_1^{-1}
 \mb{with} V = -q^{-1} E^{11} + E^{22} - q E^{33}. \label{Vmat} 
 \een

\end{document}